\documentclass{sig-alternate-2013}

\usepackage{balance}  
\usepackage{graphics} 
\usepackage{times}    
\usepackage[hyphens]{url}     
\usepackage{multirow}
\usepackage{xcolor,colortbl}
\usepackage{bchart}
\usepackage{multicol, blindtext}
\usepackage{caption}
\usepackage{subcaption}
\usepackage{amsmath}

\def\myPosBar#1{~#1 {\color[rgb]{0.17,0.43,0.70}\rule{#1cm}{5pt}} }
\def\myNegBar#1{#1 {\color[rgb]{0.84,0.11,0.09}\rule{-#1cm}{5pt}} }

\definecolor{Gray}{gray}{0.85}

\newcommand{\mytablesize}{\scriptsize}

\permission{Pre-print Version}
\conferenceinfo{WWW 2016,}{April 11--15, 2016, Montr\'eal, Qu\'ebec, Canada.} 
\copyrightetc{}

\begin{document}

\title{Who Benefits from the ``Sharing'' Economy of Airbnb?}

\numberofauthors{5} 
\author{
   \alignauthor Giovanni Quattrone\\
       \affaddr{Dept. of Geography}\\
       \affaddr{University College London, UK}\\
       \email{g.quattrone@cs.ucl.ac.uk}
   \alignauthor Davide Proserpio\\
       \affaddr{Dept. of Computer Science}\\
       \affaddr{Boston University, USA}\\
       \email{dproserp@bu.edu}
   \alignauthor Daniele Quercia\\
       \affaddr{Bell Laboratories}\\
       \affaddr{Cambridge, UK}\\
       \email{quercia@cantab.net}
\and  
   \alignauthor Licia Capra\\
       \affaddr{Dept. of Computer Science}\\
       \affaddr{University College London, UK}\\
       \email{l.capra@ucl.ac.uk}
   \alignauthor Mirco Musolesi\\
       \affaddr{Dept. of Geography}\\
       \affaddr{University College London, UK}\\
       \email{m.musolesi@ucl.ac.uk}
}

\maketitle


\begin{abstract}

Sharing economy platforms have become extremely popular in the last few years, and they have changed the way in which we commute, travel, and borrow among many other activities. Despite their popularity among consumers, such companies are poorly regulated. For example, Airbnb, one of the most successful examples of sharing economy platform, is often criticized by regulators and policy makers. While, in theory, municipalities should regulate the emergence of Airbnb through evidence-based policy making, in practice, they engage in a false dichotomy: some municipalities allow the business without imposing any regulation, while others ban it altogether. That is because there is no evidence upon which to draft policies. Here we propose to gather evidence from the Web. After crawling Airbnb data for the entire city of London, we find out where and when Airbnb listings are offered and, by matching such listing information with census and hotel data, we determine the socio-economic conditions of the areas that actually benefit from the hospitality platform. The reality is more nuanced than one would expect, and it has changed over the years.
Airbnb demand and offering have changed over time, and traditional regulations have not been able to respond to those changes. That is why, finally, we rely on our data analysis to envision regulations that are responsive to real-time demands, contributing to the emerging idea of ``algorithmic regulation''.

\end{abstract}













\section{Introduction}








In the last few years, we have seen the proliferation of sharing economy platforms. These platforms leverage information technology to empower users to share and make use of underutilized goods and services. Services covered by the sharing economy range from transportation to accommodation to finance. One of the most compelling example of the sharing economy is Airbnb, a peer-to-peer accommodation website. Airbnb defines itself as ``A social website that connects people who have space to spare with those who are looking for a place to stay''. The company, founded in 2008, grew exponentially in the past few years, and by now it lists  over 1.5 million properties, with a presence in over 190 countries and 34,000 cities. By the end of 2014, the company had more than 70M nights booked.\footnote{See: \small\url{http://www.reuters.com/article/2015/09/28/us-airbnb-growth-idUSKCN0RS2QK20150928}} 

The explosive growth of the sharing economy has led regulatory and political battles around the world. Proponents of the sharing economy argue that it will bring many benefits, including extra incomes from the users of such services, better resource allocation and utilization, and new economic activities for cities and municipalities.\footnote{Airbnb itself released several studies quantifying the positive economic impact of the company in many cities around the world. For more details see: \url{https://www.airbnb.com/economic-impact}} On the other side, detractors argue that the negative externalities generated by the sharing economy far outpace the benefits. Most of the critics denounce the sharing economy for being about economic self-interest rather than sharing, and for being predatory and exploitative. Indeed, the predatory aspect of such economy has already seen its first victims: after Uber entered the New York City market, the price of taxi medallion fell down by about 25\%,\footnote{See: \small\url{http://www.nytimes.com/2015/01/08/upshot/new-york-city-taxi-medallion-prices-keep-falling-now-down-about-25-percent.html}} and in \cite{zervas2015impact} the authors show that Airbnb entry in the state of Texas negatively impacted hotel revenue. 

Because of such negative externalities, the sharing economy and its regulation have become highly popular policy topics. Many municipal governments are attempting to impose old regulations on these new marketplaces without much thought about whether these laws apply to these companies, and without a complete understanding of the benefits and  drawbacks generated by these new services. 
Furthermore, such a debate has resulted into little academic work, as we shall see in Section~\ref{sec:related}. We aim to fill this gap by performing the 
first socio-economic analysis of Airbnb adoption. We do so by using the city of London as case study. London is particular well-suited because of its high diversity in socio-economic and geographic terms, and of its enthusiastic adoption of Airbnb  (by June 2015, London had over 14,000 Airbnb properties listed).
We show which areas benefit from Airbnb, and how the insights related to that inform policy making. More specifically, we make two main contributions:

\begin{itemize}
\item We crawl Airbnb data in London from 2012 to 2015 and study the adoption of the platform across  the UK census areas in the city (Section~\ref{sec:datasets}).

\item We analyze such data (Section~\ref{sec:method}) and contrast the socio-economic conditions of the areas that benefit from Airbnb to those of the areas that do not (Section~\ref{sec:results}). 
\end{itemize}

We then conclude by putting forward five recommendations on how Airbnb might be regulated based on our insights (Section~\ref{sec:discussion}).

\section{Related Work}
\label{sec:related}
Our work relates to the growing literature on the regulation of the sharing economy. Research in these area comes from many disciplines, from law to economy to policy. 
In \cite{edelman2015efficiencies}, the authors, after enumerating the efficiencies that the sharing economy provides for both service providers and consumers, discuss regulation and policies for such software platforms. They suggest the need to adapt law and regulations to allow those platforms to operate legally. This will ensure that service providers, users and third parties are adequately protected from any harm that may arise. Of the same opinion are the authors in \cite{koopman2014sharing}. They  argue that when market circumstances change dramatically -- or when new technology or competition alleviates the need for regulation -- then public policy should accordingly evolve. Einav {\em et al.}~\cite{einav2015peer} provide a discussion about licensing, employment regulation, data, and privacy regulation of the sharing economy. They do so by considering the current regulations adopted by a few municipalities, and discussing the pros and cons. In \cite{miller2015first}, the author critiques the  existing regulation of Airbnb. \cite{zale2015sharing} presents a taxonomy of ``sharing'', including formality and gratuity, and examines doctrinal responses to sharing situations. \cite{rogers2015social} compares Uber's efficiencies with its regulatory arbitrage.
\cite{ranchordas2015does} analyzes the challenges of regulating the sharing economy from an ``innovation law perspective'' by arguing that these innovations should not be stifled by regulation, but should also not be left totally unregulated. \cite{cohenself} argues for self-regulatory approaches and reallocation of regulatory responsibility to parties other than the government. Finally, \cite{Ikkala}  studies how financial incentives are mediated by hospitality and sociability in Airbnb. 

While the above works do an excellent job in defining the bases upon which the sharing economy should be regulated, none of them does so upon empirical evidence of what the sharing economy really is, how it has been adopted, and who benefits from it. By contrast, this work argues for evidence-informed policy making, and provides answers to the above questions by empirically investigating Airbnb adoption.


\section{Overview}
\label{sec:overview}



\begin{figure*}[tb]
\centering
  \begin{subfigure}{.3\textwidth}
    \centering
    \includegraphics[width=0.98\linewidth]{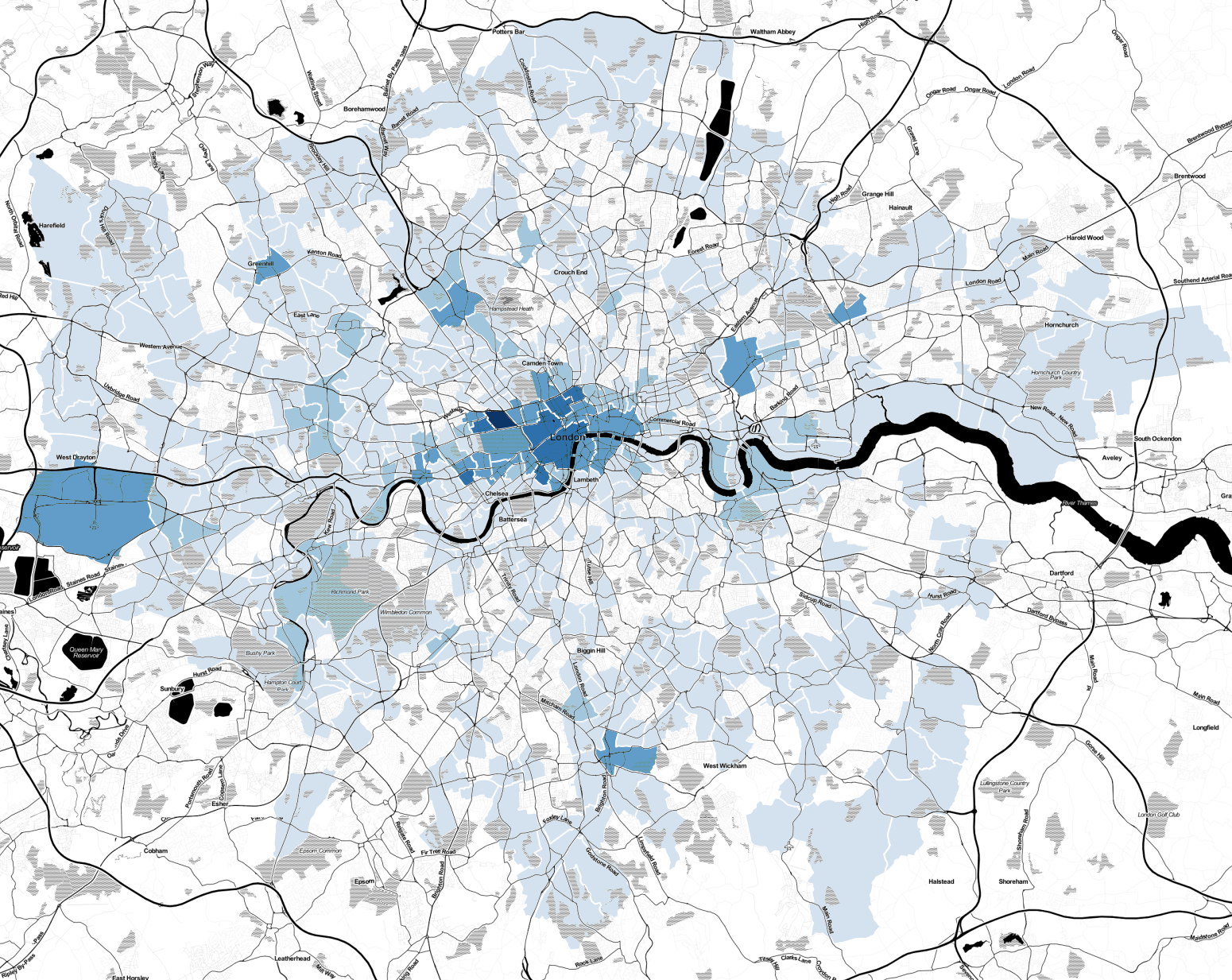}
    \caption{Hotels}
    \label{fig:hotels}
  \end{subfigure}%
  \begin{subfigure}{.3\textwidth}
    \centering
    \includegraphics[width=0.98\linewidth]{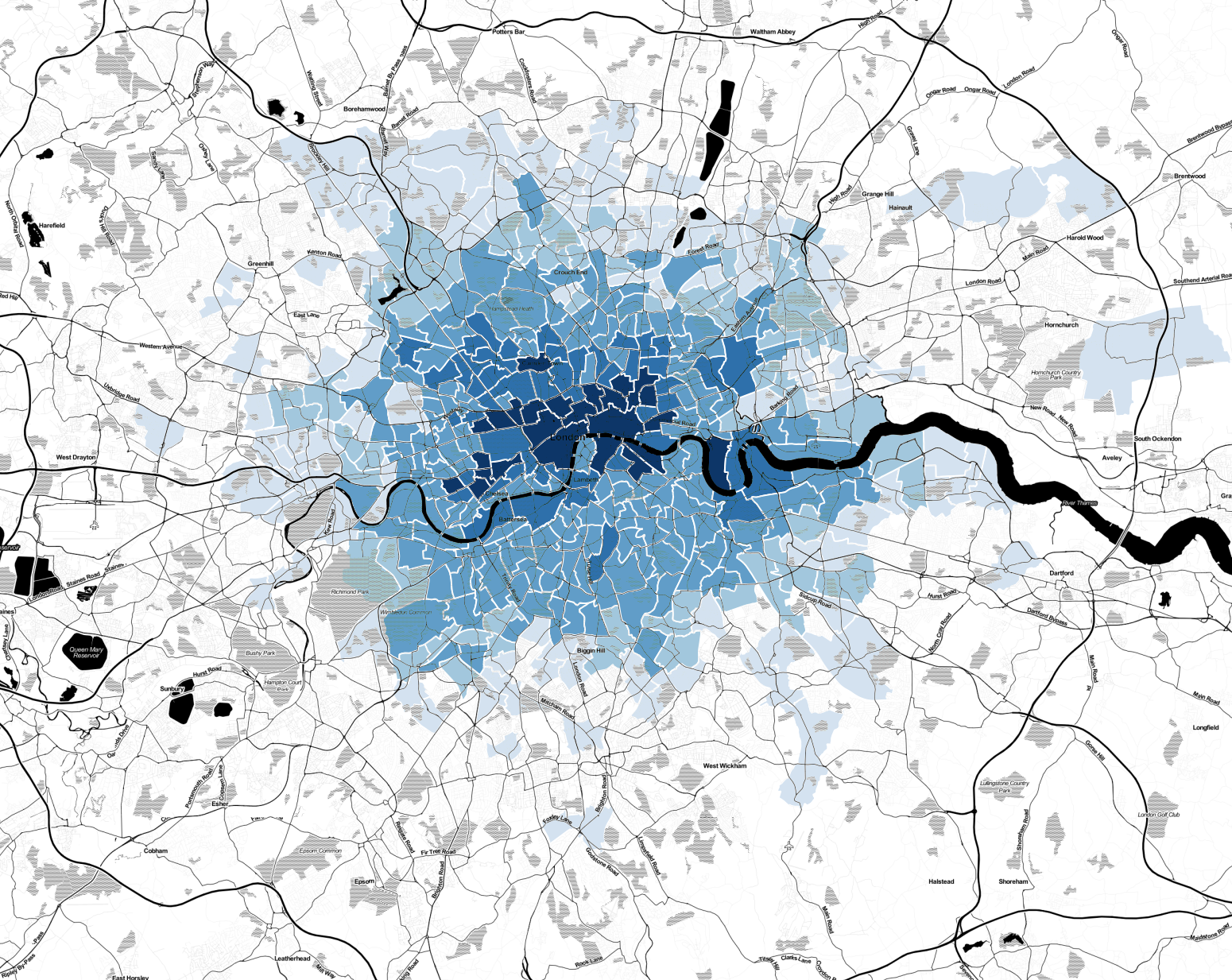}
    \caption{Airbnb houses}
    \label{fig:airbnb-houses}
  \end{subfigure}
  \begin{subfigure}{.3\textwidth}
    \centering
    \includegraphics[width=0.98\linewidth]{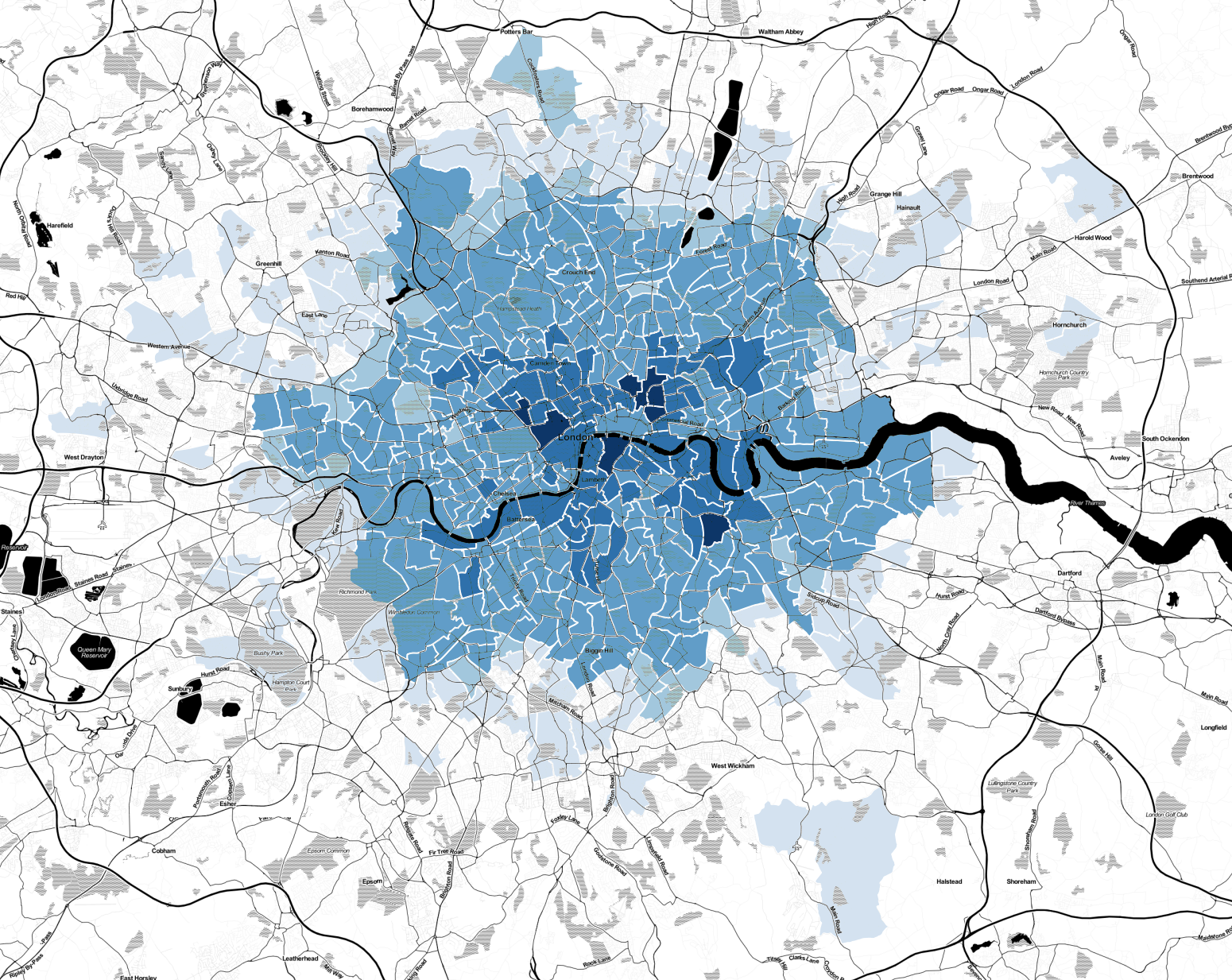}
    \caption{Airbnb rooms}
    \label{fig:airbnb-rooms}
  \end{subfigure}%
  \begin{subfigure}{.08\textwidth}
    \centering
    \includegraphics[width=0.98\linewidth]{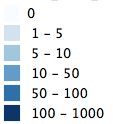}
  \end{subfigure}%
  \caption{Heat maps of the number of hotels, Airbnb houses, and Airbnb rooms in each London ward. The darker the ward, the higher the number. The legend reflects the actual (not normalized) numbers, which are thus comparable across the three maps.}
  \label{fig:map-hotels-airbnbs}
\end{figure*}

Where are Airbnb listings  located? This is one of the most frequently asked questions by municipalities, hoteliers and travellers.  To start answering it, we crawled extensive data about Airbnb properties (from 2012 to 2015) and hotels for the city of London (which the next section will describe in detail), and we simply map the presence of hotels and Airbnb listings in the city.
A clear distinction that the Airbnb website makes is between \textit{entire home/apartment} (case where the whole home/apartment is rented) and \textit{private room} (case where only a private room is rented and all the other spaces of the house are shared with others). Given that distinction, we separately map the offering of  Airbnb houses (Figure~\ref{fig:airbnb-houses}) and Airbnb rooms (Figure~\ref{fig:airbnb-rooms}), and contrast them to the offering of hotels (Figure~\ref{fig:hotels}).
Figure~\ref{fig:map-hotels-airbnbs} shows that hotels have spotty coverage throughout the city of London, and they are mostly concentrated in the center and near the main airport (Heathrow) on the west side. Airbnb houses have a heavy presence in the city center (like hotels), but they also reach adjacent areas up to around 10 miles from the center. Airbnb rooms massively cover -- almost uniformly -- the greatest part of the city of London instead,  including suburban areas.

%

To go beyond visual inspection, we compute the overlap between Airbnb adoption and hotel adoption. Since each area can be covered at various levels of strengths by Airbnb and hotels, we adopt the fuzzy logic functions. Specifically let $bnb$ and $hotel$ be two fuzzy sets such that $bnb_i \in [0,1]$ and $hotel_i \in [0,1]$ denote, respectively, the strength of Airbnb's offering in area $i$ and of hotels'.  The strength is zero if Airbnb listings (hotels) are totally absent from area $i$,  is one if they show maximum presence (with respect to the entire dataset), and, otherwise, assumes intermediate values proportional to the presence. Upon those two sets, we compute the ratio of areas covered by Airbnb that are  also  covered by hotels, and the ratio of areas covered by hotels that are also covered by Airbnb as follows:
   \begin{equation} \label{eq:overlap}
   \begin{aligned}
      hotel\_in\_bnb\_areas = \frac{| bnb \cap hotel |}{| bnb |} \\
      bnb\_in\_hotel\_areas = \frac{| bnb \cap hotel |}{| hotel |} \\
   \end{aligned}
   \end{equation}
Where the intersection $(bnb \cap hotel)$ of two fuzzy sets  is defined by $(bnb \cap hotel)_i = min \{ bnb_i, hotel_i \}$, and the cardinality of a fuzzy set $bnb$ is defined by $|bnb| = \sum_{i}{bnb_i}$. 
We compute these ratios for every year, from 2012 to 2015, for the two Airbnb listings categories:  rooms and  houses (Table~\ref{tab:overlapping}). Airbnb properties (especially rooms) tend to be located in areas where there are hotels. That has been true over the years and, from 2012 to 2015, has increased  for Airbnb houses as well (specifically, by 7\%).  On the contrary, hotels do not tend to be in areas where there are Airbnb properties. Therefore, we can safely conclude that Airbnb listings cover a much broader city area than what hotels do.

\begin{table}[t!]
\mytablesize
\centering
\begin{tabular}{lcc}
\hline 
\rowcolor{Gray} 
        & \multicolumn{2}{c}{\em Airbnb Rooms} \\
\hline
   Year & $hotel\_in\_bnb\_areas$ & $bnb\_in\_hotel\_areas$ \\
\hline \hline
2012 & 0.14  & 0.64  \\
2013 & 0.14  & 0.67  \\
2014 & 0.12  & 0.71  \\
2015 & 0.12  & 0.71  \\
\hline 
\rowcolor{Gray} 
        & \multicolumn{2}{c}{\em Airbnb Houses} \\
\hline
   Year & $hotel\_in\_bnb\_areas$ & $bnb\_in\_hotel\_areas$ \\
\hline \hline
2012 & 0.24  & 0.64  \\
2013 & 0.23  & 0.64 \\
2014 & 0.24  & 0.64 \\
2015 & 0.24  & 0.63 \\
\hline \hline
\end{tabular}
\caption{Fraction of London areas that have hotels and Airbnb properties (rooms \emph{vs.} houses).}
\label{tab:overlapping}    
\end{table}

Since the spatio-temporal dynamics behind Airbnb are quite unique (and definitely different than those behind hotels), we set out to study it in detail and answer four main questions:

\begin{itemize}
\item[{\bf RQ1}] -- What are the main socio-economic characteristics of areas with Airbnb listings?

\item[{\bf RQ2}] -- Are all types of listings equal? Is there any difference between, for example, Airbnb listings of rooms and those of entire houses? 

\item[{\bf RQ3}] -- What is the temporal evolution of Airbnb listings? 

\item[{\bf RQ4}] -- Where do Airbnb customers actually go? That is, what are the main socio-economic characteristics of areas where Airbnb customers go?
\end{itemize}




\section{Datasets and Metrics}
\label{sec:datasets}


To answer those questions, we need to collect information from various data sources. On one hand, we need detailed records of Airbnb properties; on the other hand, we need to collect socio-economic data and derive neighborhood metrics from it.

\begin{table*}[!t]
   \mytablesize
   \centering
   \begin{tabular}{ c | l | l | m{9cm}} 
      \hline 
      \rowcolor{Gray} 
      \multicolumn{1}{c}{\em Category} & \multicolumn{1}{|c}{\em Metric} & \multicolumn{1}{|c}{\em Source} & \multicolumn{1}{|c}{\em Description}\\
      \hline
      \multirow{2}{*}{Airbnb} 
         & $bnb\_\mathit{offering}$ & Airbnb website & Number of Airbnb properties per $km^2$\\
         & $bnb\_demand$ & Airbnb website & Number of Airbnb reviews per $km^2$\\
      \hline
      Hotel
         & $hotel\_\mathit{offering}$ & Ordnance Survey & Number of hotels per $km^2$ \\
      \hline
      \multirow{3}{*}{Attractiveness} 
         & $foursquare$ & Foursquare & Number of Foursquare check-ins per $km^2$ \\
         & $transport$ & Census & Score for accessibility to public transportation\\
         & $attractions$ & Ordnance Survey & Number of attractions and entertainment places \\
      \hline
      \multirow{7}{*}{Demographic} 
         & $young$ & Census &  Number of people aged between 20 and 34 years per $km^2$  \\
         & $income$ & IMD from Census & Score for income \\
         & $employment$ & Census &  Ratio of the number of employees over the area's population\\
         & $ethnical\_mixed$ & Census &  Score for ethnic diversity \\
         & $bohemian$ & Census &  Fraction of residents employed in arts, entertainment, and recreation \\
         & $melting\_pot$ & Census &  Percentage of non-UK born residents\\
         & $education$ & Census &  Percentage of residents with MSc+\\
      \hline
      \multirow{7}{*}{Housing} 
         & $living$ & IMD from Census & Score for living environment conditions\\
         & $green\_space$ & Census &  Percentage of green space over the total area's surface \\
         & $top\_house\_price$ & Census &  Percentage of dwellings in council tax band F-H (band of the highest median house price)\\
         & $houses\_vs\_flats$ & Census &  Percentage of houses over houses plus flats \\
         & $owned\_vs\_rented$ & Census &  Percentage of owned properties \\
         & $house\_price$ & Census &  Median house price\\ 
         & $sold\_houses$ & Census &  Number of properties sold per $km^2$ \\      
   \hline \hline           
   \end{tabular}
\caption{Description of the variables used in our analyses.}
\label{tab:indep-vars}
\end{table*}


\subsection{Airbnb Data}
\label{sec:Airbnb_metrics}

We have periodically collected, since mid 2012, consumer-facing information from \url{airbnb.com} on the complete set of users who had listed their properties in the city of London for rental on Airbnb. We refer to these users as \textit{hosts}, and their properties as their \textit{listings}. Each host is associated with a set of attributes including a photo, a personal statement, their listings, guest reviews of their properties, and Airbnb-certified contact information. 
Similarly, each listing displays attributes including location, price, a brief textual description, photos, capacity, availability, check-in and check-out times, cleaning fees, and security deposits. 

Our collected dataset contains detailed information on 14,639 distinct London hosts,  17,825 distinct London listings, and 220,075 guest reviews spanning a period from March 2012 to June 2015. From this data we measure:

\begin{description}
\item \emph{Airbnb offering per area (bnb\_offering):} the ratio between the number of Airbnb listings registered in a given London area over the surface of the same area in square kilometers. We have also considered two types of normalization other than surface -- number of inhabitants and number of dwellings. For all the three types, the results are comparable. 

\item \emph{Airbnb demand per area (bnb\_demand):} the total number of Airbnb reviews registered in a certain area of London over the size of the area in square kilometers. We use reviews as a proxy for demand, not least because it has been shown that people leave reviews after staying at a place more than 70\% of the times~\cite{fradkin2015bias}. 
\end{description}




\subsection{Socio-economic Conditions}

We used two different data sets that reflect socio-economic conditions of London areas.

~

\subsubsection{Census Data}

We gather the 2011 official UK census data\footnote{See: \small\url{http://data.london.gov.uk/dataset/ward-profiles-and-atlas}} containing demographic information about small areas defined by the UK Government and known as {\em wards}. This includes the population density of the area, how many young people live there, the number of educated people, as well proxies concerning how pleasant a particular area is to live in (e.g., the percentage of green space). From this dataset, we also collect housing information. This includes the number of flats and houses present in an area, the number of properties sold, the number of dwellings that are owned rather than rented, and the median house price. This information is useful to have an accurate picture of the type of housing available in each London area, as well as the fluidity of the housing market there. Most of those metrics have been widely used. By contrast, a few have been used in a limited number of papers and need to be illustrated:

\begin{description}
\item {\em Diversity of Ethnic Groups} ($ethnical\_mixed$). The idea for this diversity index was taken from Chris von Csefalvay's data blog~\cite{Csefalvay2014}. In the blog, the author describes a method of measuring diversity in England and Wales with a metric taken from mathematical ecology. This metric is calculated as the Gini-Simpson diversity index\footnote{The Simpson diversity index is a measure that reflects how many different entries there are in a data set and the value is maximized when all entries are equally high~\cite{simpson1949measurement}.} of the ethnic groups living in each area. The census data contains five different categories of ethnicity (number of white, black, Asian, mixed and other individuals in an area). These five categories were used to calculate the Gini-Simpson index. This index represents the probability that two individuals chosen at random from an area are of a different ethnicity (high values are associated with multi-ethnic areas).
\item {\em Bohemian Index} ($bohemian$). We start from the work of Richard Florida~\cite{florida2008there} on the effect of the bohemian, artistic and gay population on regional house prices. The author found that a newly derived ``Bohemian-Gay Index'' has a substantial effect on house prices. We can thus hypothesize that a similar metric may have an interesting effect on the number or price of Airbnb offerings. Unfortunately, since gender is not part of the  UK census information, we are not able to recreate this metric. We therefore followed the same approach adopted by Nick Clifton \cite{clifton2008creative} that analyzed the creative class in the UK instead. By following Florida's work, Clifton computed a cultural metric (the Bohemian Index) by only using the data made available in the UK census. This metric is defined as the fraction of people employed in arts, entertainment and recreation.
\item {\em Melting Pot Index} ($melting\_pot$). This is the second metric used by Nick Clifton~\cite{clifton2008creative} to describe the creative class in the UK and is the number of people born outside the UK divided by the total number of people in the area.
\end{description}

\medskip

\subsubsection{IMD Score}

We also collect  the UK Index of Multiple Deprivation (IMD) data\footnote{See: \small\url{https://www.gov.uk/government/uploads/system/uploads/attachment_data/file/6871/1871208.pdf}} available at the level of small census areas known as Lower-layer Super Output Areas (LSOAs). LSOAs are defined to roughly include always the same number of inhabitants (around 1,500).\footnote{See: \small\url{https://www.gov.uk/government/statistics/english-indices-of-deprivation-2010}} IMD is a composite score, comprising seven distinct domains: {\em (i)}~income, {\em (ii)}~employment, {\em (iii)}~health, {\em (iv)}~education, {\em (v)}~barrier to housing and services, {\em (vi)}~crime, and {\em (vii)}~living environment. For the purpose of our study, we collected the values of two indexes, called income and living environment, as we hypothesize that these two factors, jointed with the ones collected with the census data, may have an impact on the number and type of Airbnb offerings.

\medskip
\subsection{Attractiveness}

A traditional metric often used to describe London areas is transportation accessibility ($transport$): the higher the value, the more accessible the area by public transport. This metric is ready available from the UK Census. To capture more nuanced facets of attractiveness of London areas other than transport accessibility, we compute three further metrics from two other data sets.

\subsubsection{Foursquare}
Foursquare has been launched in 2009 and it is one of the most popular location-based social networking website.\footnote{See: \small\url{https://foursquare.com/about}} Using Foursquare, registered users that visit a location can ``check-in'' on the application to share their real-time location with friends. In December 2013, Foursquare surpassed 45 million registered users and currently male and female users are equally represented.\footnote{See: \small\url{https://en.wikipedia.org/wiki/Foursquare}} Janne Lindqvist \emph{et al.}  studied why people check-in and found that individuals tend to use Foursquare to see where they have been in the past and ultimately curate their own location history~\cite{lindqvist2011m}. For this reason, we hypothesize that, in cities where Foursquare has high penetration such as London, the number of Foursquare check-ins may be considered as an approximate measure of the attractiveness of  areas (i.e.,  areas where city dwellers prefer to visit and spend time in). We use the official Foursquare API to crawl Foursquare check-ins.\footnote{See: \small\url{https://api.foursquare.com}} We perform this step between 04/03/2014 and \break 08/04/2014, resulting in the collection of 26,344,115 users check-ins in the whole London metropolitan area. We then compute our first measure of area attractiveness as the number of Foursquare check-ins in a specific area over the area's surface in square kilometers. We denote this variable as $foursquare$.  

\subsubsection{Ordnance Survey}
%
%
%
%

Ordnance Survey~\footnote{See: \small\url{www.ordnancesurvey.co.uk}} is the national mapping agency for Great Britain. OS mapping is usually classified as the more detailed mapping of the country and covers not only roads but also millions of Point of Interests (POIs) of varying nature, from restaurants to hospitals and hotels. Ordnance survey data is freely available.~\footnote{See: \small\url{www.ordnancesurvey.co.uk/resources}} We downloaded the data in July 2015, collecting 513,786 POIs in the whole metropolitan London area. For the purpose of this study, we considered the number of Ordnance Survey POIs that fall under one of the categories of ``eating and drinking'', ``attractions'', ``retail'', ``sports and entertainment'' -- to capture London areas that are covered by attractions -- normalized by the size of the area in square kilometers. We denote this variable as $attractions$.


\subsection{Hotel Data}

To study whether Airbnb offerings are located in areas with presence of traditional forms of accommodation, we consider the number of Ordnance Survey POIs that fall under one of the categories of ``hotels'', ``motels'', ``country houses and inns'' normalized by the size of the area in square kilometers. We denote this variable as $hotel\_offering$.  
%
%
%
%
Table~\ref{tab:indep-vars} lists all the metrics that we have computed so far, and that we use next.

\section{Method}
\label{sec:method}

This section describes the method we have developed to answer the  four questions we put forward in the final part of Section~\ref{sec:overview}.

\subsection{Unit of Analysis}

The goal of this work is to measure the number and the type of Airbnb offerings in different areas of a city (London, in our case) and study their relationship between offering and neighborhood socio-economic conditions. 

\begin{figure}[t]
    \centering
    \includegraphics[width=0.98\linewidth]{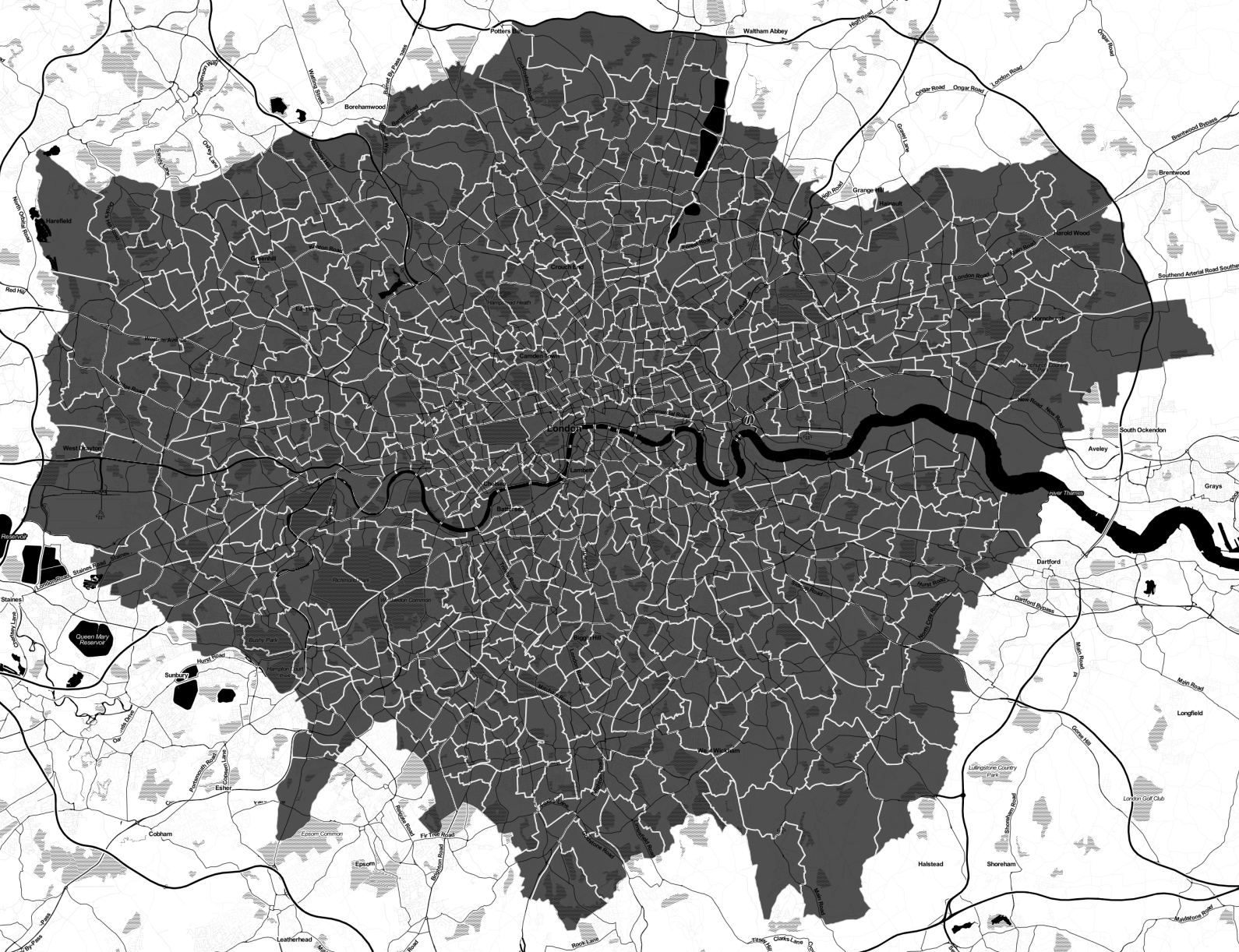}
    \caption{London Wards.}
    \label{fig:wards}
\end{figure}

To do so, we need to define a {\em spatial unit} of analysis that is representative of the different London areas. We therefore choose a spatial unit called {\em ward}. Using official geographic definitions of wards in the UK,\footnote{See: \small\url{https://geoportal.statistics.gov.uk/Docs/Boundaries/Wards_(E+W)_2011_Boundaries_(Full_Extent).zip}} we computed 625 wards for Greater London illustrated in Figure~\ref{fig:wards}. Although we are aware that wards might not be completely homogeneous in terms of their characteristics, the size of wards allows for the collection of a statistically significant number of data points, which is not possible to obtain by using smaller geographic units such as LSOAs. From now on, for the sake of simplicity, we will refer to wards as ``areas''. Whenever  data was available at a different level of granularity, we aggregated it to have the information at ward level. This was the case, for example, of IMD data, which was available at level of LSOA; in such case, we computed the IMD score of a ward as the average of the IMD scores of the ward's LSOAs. Very little information was lost during such an aggregation step, as IMD scores for LSOAs in the same ward are very consistent (the standard deviation is less than the corresponding average value, for all wards).

In terms of {\em temporal unit} of analysis, we must concede that there is a four-year gap between the UK census data of 2011 and the other sets of data that refer to 2014/2015.  However, even in the presence of this gap,  a cross-comparison of all sets is still possible. That is because the collection of census data is conducted every 10 years in the UK and, as such, the census indicators are bound to remain unchanged in a 4-year time window.

\subsection{Approach}

The aim of this paper is to give insights about who benefits from the economy generated by Airbnb. As a first step, we study which of the socio-economic factors associated with the London areas are significantly correlated with Airbnb offering. We use a linear regression model in the form of Ordinary Least Squares (OLS):
   \begin{equation} \label{eq:OLS}
      Y_i = \beta_0 + \beta_1 X_i + \epsilon_i \ \ ,
   \end{equation}
where $Y_i$ is one of the first three metrics in Table~\ref{tab:indep-vars} (i.e., $bnb\_offering$, $bnb\_demand$, and $hotel\_offering$ for area $i$), and $X_i$ is the set of the remaining metrics in the table, which reflect the socio-economic conditions of area $i$. 

Since we are dealing with geographic data, for each produced model, we test for the presence of spatial autocorrelation. This is the tendency for measurements located close to each other to be correlated, a property that generally holds for variables observed across geographic spaces~\cite{legendre1993spatial}. We test our OLS models for spatial auto-correlation by computing the Moran's test\footnote{The Moran's test is a measure of spatial autocorrelation developed by Patrick Alfred Pierce Moran~\cite{moran1950notes}. Values range from $-1$ (indicating perfect spatial dispersion) to $+1$ (perfect spatial auto-correlation). A zero value indicates a random spatial pattern.} on the residuals $\epsilon_i$. 

Finally, since most of our metrics are skewed and therefore do not conform with the normality assumption of the variance, we improve the normality of such variables by applying a log transformation. Further, since our metrics are on very different scales, we standardize them by computing their $z$-scores. This transformation enables us to compare $\beta$ scores that are from different distributions.


\section{Results}
\label{sec:results}

This section is subdivided in two parts: in the first, we present some preliminary results coming out from a cross-correlation analysis performed on the adopted metrics; in the second,  we describe the results we have collected to answer our four research questions.

\subsection{Preliminary Analysis}
\label{sec:preliminary_results}

\begin{figure}[t]
    \centering
    \includegraphics[width=0.98\linewidth]{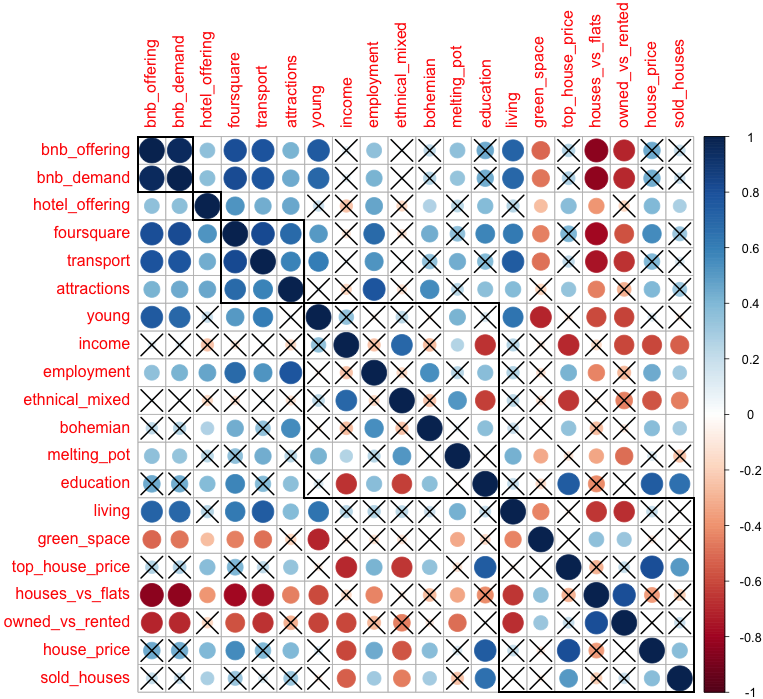}
    \caption{Pearson cross-correlation matrix of the metrics in Table~\ref{tab:indep-vars} with significance levels (crossed circles indicate $p$-value $>0.01$). The cross-correlations are grouped according to the classification in the table.}
    \label{fig:crosscorr}
\end{figure}

To know which of our variables in Table~\ref{tab:indep-vars} are correlated with each other, we compute the cross-correlation matrix (Figure~\ref{fig:crosscorr}). Take the first row. It shows which variables are correlated with Airbnb offering. We see that Airbnb listings tend to be in areas that are attractive and accessible by public transport, and that have residents who are young, employed, and born outside UK. By contrast, Airbnb listings tend not to be in areas where there are more houses than flats and where there are more owned properties than rented ones (these areas are likely to be suburban areas). Similar results can be found when looking at the second row of Figure~\ref{fig:crosscorr}, which looks at Airbnb demand.


These initial results suggest that our conjecture that specific neighborhood socio-economic conditions are related to Airbnb demand and offering is well-grounded. To now go into the details, we perform a regression analysis.  Since some of our independent variables exhibit levels of cross-correlations (Figure~\ref{fig:crosscorr}), we expect that not all the variables that are now correlated with Airbnb offering and demand will maintain the same significance levels in the next regression analysis.




\subsection{RQ1. Socio-economic Conditions}
\label{sec:conditions}

After mapping the Airbnb listings (Figure~\ref{fig:map-hotels-airbnbs}), we have observed that the offering seems to be highly correlated with the distance from the city center. For this reason, we regress the cumulative values of Airbnb offering $bnb\mathit{\_offering}$ per London area against the socio-economic variables described in Section~\ref{sec:datasets} plus distance from the city center. A previous study has found that London has 10 different polis~ \cite{batty10}. In this work, we thus computed $distance$, which is the Euclidean distance from the geographic center point of each ward to the geographic center point of each of the 10 polis. We then used the shortest distance as our ``distance from the center'' factor, and tested the hypothesis that the closer to the poly-center, the higher the offering and the demand of Airbnb.

The estimates of the regression model are reported in Table~\ref{tab:AirA}. Indeed, in the second row, one sees that the farther the distance from the center, the lower the number of listings. Also, Airbnb properties are, again, associated with attractive and well-to-do areas with young and tech-savvy residents. 

\begin{table}[t]
    \mytablesize
    \tabcolsep 3pt
    \centering
	 \begin{tabular}{llrlrl}
     \hline \hline
                     & {\em ~~Indep. var}      & $p$-val & ~~$\beta$ \\
     \hline \hline
     Hotel           & $hotel\_\mathit{offering}$       &     & \myNegBar{-0.02} \\
     Geography       & $distance$      & *** & \myNegBar{-0.25} \\
     Attractiveness  & $foursquare$         & **  & \myPosBar{0.14}  \\
                     & $transport$    &     & \myNegBar{-0.05} \\
                     & $attractions$              &     & \myPosBar{0.02}  \\
     Demographics    & $young$       & *** & \myPosBar{0.40}  \\
                     & $income$      & *** & \myNegBar{-0.16} \\
                     & $employment$    &     & \myPosBar{0.00}  \\
                     & $ethnical\_mixed$ & **  & \myPosBar{0.09}  \\                     
                     & $bohemian$    &     & \myNegBar{-0.01} \\
                     & $melting\_pot$       & *   & \myNegBar{-0.07} \\
                     & $education$   &     & \myNegBar{-0.01} \\
     Housing         & $living$ & .   & \myPosBar{0.05}  \\
                     & $green\_space$        &     & \myPosBar{0.03}  \\
                     & $top\_house\_price$     &     & \myPosBar{0.03}  \\
                     & $houses\_vs\_flats$        & .   & \myNegBar{-0.10} \\
                     & $owned\_vs\_rented$         & *** & \myNegBar{-0.23} \\
                     & $house\_price$       & *** & \myPosBar{0.15}  \\
                     & $sold\_houses$        & **  & \myPosBar{0.07}  \\
     \hline \hline
                     & {Adjusted R-squared} &  & ~0.90 \\ 
                     & {Moran's test}       &  & ~0.03 \\ 
	 \end{tabular}
	 \caption{Analysis of Airbnb offering. Blue (resp., red) bars reflect positive (resp., negative) slope coefficients.}
    \label{tab:AirA}
\end{table}

\subsection{RQ2. Airbnb Rooms vs. Houses}
\label{sec:rooms-vs-houses}


So far we have provided evidence that, when treating Airbnb listings homogeneously, properties are
more likely to be concentrated in tech-savvy and well-to-do areas with young renters. In practice, Airbnb listings are very different among them though. A clear distinction that the website makes is between \textit{entire homes/ \break apartments} and \textit{private rooms}.
Therefore, in this section, we repeat the above analysis by separating Airbnb listings into those two categories (Table~\ref{tab:AirA_rooms_vs_houses}). 
We observe significant differences:  Airbnb \emph{rooms} tend to be offered in areas with highly-educated non-UK born renters, while  \emph{homes} tend to be offered in areas with owners of high-end homes in terms of house price. 



\begin{table}[t]
    \mytablesize
    \tabcolsep 3pt
    \centering
	 \begin{tabular}{llrlrl}
     \hline
     \rowcolor{Gray} 
     & & \multicolumn{2}{c}{\em Airbnb Rooms} & \multicolumn{2}{c}{\em Airbnb Houses} \\
     \hline
                     & {\em ~~Indep. var}      & $p$-val & ~~$\beta$      & $p$-val & ~~$\beta$ \\
     \hline \hline
     Hotel           & $hotel\_\mathit{offering}$        &     & \myNegBar{-0.02} & .   & \myNegBar{-0.04} \\ 
     Geography       & $distance$       & *** & \myNegBar{-0.33} & *** & \myNegBar{-0.16} \\
     Attractiveness  & $foursquare$          & **  & \myPosBar{0.15}  & *   & \myPosBar{0.12} \\ 
                     & $transport$     & .   & \myNegBar{-0.07} &     & \myNegBar{-0.04} \\
                     & $attractions$               &     & \myPosBar{0.03}  &     & \myPosBar{0.01} \\ 
     Demographics    & $young$        & *** & \myPosBar{0.44}  & *** & \myPosBar{0.29} \\ 
                     & $income$       & **  & \myNegBar{-0.13} & *   & \myNegBar{-0.12} \\ 
                     & $employment$     &     & \myNegBar{-0.06} &     & \myPosBar{0.03} \\ 
                     & $ethnical\_mixed$ & *** & \myPosBar{0.13}  & **  & \myPosBar{0.10} \\ 
                     & $bohemian$     &     & \myNegBar{-0.01} &     & \myPosBar{0.00} \\ 
                     & $melting\_pot$        & *** & \myNegBar{-0.11} &     & \myNegBar{-0.05} \\ 
                     & $education$    & *   & \myPosBar{0.10}  &     & \myPosBar{0.01} \\ 
     Housing         & $living$  &     & \myPosBar{0.03}  & *   & \myPosBar{0.08} \\ 
                     & $green\_space$         &     & \myPosBar{0.03}  &     & \myPosBar{0.01} \\ 
                     & $top\_house\_price$      & .   & \myPosBar{0.08}  &     & \myPosBar{0.05} \\ 
                     & $houses\_vs\_flats$         &     & \myPosBar{0.00}  & *   & \myNegBar{-0.14} \\ 
                     & $owned\_vs\_rented$          & *** & \myNegBar{-0.27} & *** & \myNegBar{-0.25} \\ 
                     & $house\_price$        &     & \myNegBar{-0.01} & *** & \myPosBar{0.22} \\ 
                     & $sold\_houses$         & .   & \myPosBar{0.05}  & **  & \myPosBar{0.09} \\ 
     \hline \hline
                     & {Adjusted R-squared} &  & ~0.87            &     & ~0.86 \\ 
                     & {Moran's test}       &  & ~0.02            &     & ~0.03 \\ 
	 \end{tabular}
	 \caption{Analysis of Airbnb offering by category (rooms \emph{vs.} houses).}
    \label{tab:AirA_rooms_vs_houses}
\end{table}

\subsection{RQ3. Temporal Adoption}
\label{sec:temporal-adoption}

Since our Airbnb data unfolds over four years, we are able to study its temporal characteristics. To this end, we regress the number of Airbnb listings that appear every year since 2012 (year in which Airbnb first entered the London market) against our set of socio-economic metrics. By doing so, we are able to undercover how Airbnb offering evolved over time:  which characteristics consistently explain Airbnb growth \emph{vs.} which ones change over time instead. In Table~\ref{tab:AirA_temporal}, we report the estimates obtained for the four years, from $2012$ to  $2015$.

\begin{table*}[t]
    \mytablesize
    \tabcolsep 3pt
    \centering
	 \begin{tabular}{llrlrlrlrl}
     \hline
     \rowcolor{Gray} 
     & & \multicolumn{2}{c}{\em 2012} & \multicolumn{2}{c}{\em 2013} & \multicolumn{2}{c}{\em 2014} & \multicolumn{2}{c}{\em 2015} \\
     \hline
                     & {\em ~~Indep. var}      & $p$-val & ~~$\beta$      & $p$-val & ~~$\beta$     & $p$-val & ~~$\beta$    & $p$-val & ~~$\beta$ \\
     \hline \hline
     Hotel           & $hotel\_\mathit{offering}$        &     & \myPosBar{0.00}  &     & \myNegBar{-0.02} & **  & \myNegBar{-0.09} & .   & \myNegBar{-0.08}\\
     Geography       & $distance$       & *** & \myNegBar{-0.32} & *** & \myNegBar{-0.19} & *   & \myNegBar{-0.12} & *   & \myNegBar{-0.13}\\
     Attractiveness  & $foursquare$          & *** & \myPosBar{0.20}  & .   & \myPosBar{0.12}  &     & \myNegBar{-0.06} &     & \myNegBar{-0.11}\\
                     & $transport$     &     & \myNegBar{-0.07} & .   & \myNegBar{-0.09} &     & \myPosBar{0.05}  &     & \myPosBar{0.02}\\
                     & $attractions$               &     & \myPosBar{0.06}  &     & \myPosBar{0.04}  &     & \myNegBar{-0.04} &     & \myNegBar{-0.01}\\
     Demographics    & $young$        & *** & \myPosBar{0.27}  & *** & \myPosBar{0.32}  & *** & \myPosBar{0.31}  & **  & \myPosBar{0.24}\\
                     & $income$       &     & \myPosBar{0.05}  & *** & \myNegBar{-0.21} & *** & \myNegBar{-0.42} & *** & \myNegBar{-0.62}\\
                     & $employment$     & **  & \myNegBar{-0.12} &     & \myNegBar{-0.02} &     & \myPosBar{0.07}  &     & \myPosBar{0.06}\\
                     & $ethnical\_mixed$ & *** & \myPosBar{0.17}  & **  & \myPosBar{0.12}  &     & \myPosBar{0.01}  &     & \myPosBar{0.02}\\
                     & $bohemian$     &     & \myNegBar{-0.03} &     & \myNegBar{-0.01} &     & \myPosBar{0.00}  &     & \myNegBar{-0.01}\\
                     & $melting\_pot$        & **  & \myNegBar{-0.09} & **  & \myNegBar{-0.12} &     & \myNegBar{-0.03} &     & \myNegBar{-0.02}\\
                     & $education$    &     & \myPosBar{0.07}  &     & \myPosBar{0.05}  &     & \myNegBar{-0.10} & **  & \myNegBar{-0.25}\\
     Housing         & $living$  & .   & \myPosBar{0.06}  &     & \myPosBar{0.03}  &     & \myPosBar{0.03}  &     & \myNegBar{-0.01}\\
                     & $green\_space$         &     & \myNegBar{-0.02} &     & \myPosBar{0.01}  &     & \myPosBar{0.01}  &     & \myPosBar{0.00}\\
                     & $top\_house\_price$      &     & \myPosBar{0.03}  & *   & \myPosBar{0.13}  & **  & \myPosBar{0.21}  &     & \myPosBar{0.07}\\
                     & $houses\_vs\_flats$         &     & \myNegBar{-0.07} & .   & \myNegBar{-0.13} &     & \myPosBar{0.02}  &     & \myNegBar{-0.15}\\
                     & $owned\_vs\_rented$          &     & \myNegBar{-0.11} & *** & \myNegBar{-0.32} & *** & \myNegBar{-0.64} & *** & \myNegBar{-0.59}\\
                     & $house\_price$        & *** & \myPosBar{0.20}  & *   & \myPosBar{0.13}  &     & 0\myPosBar{.04}  &     & \myPosBar{0.04}\\
                     & $sold\_houses$         & .   & \myPosBar{0.05}  &     & \myPosBar{0.03}  & *** & \myPosBar{0.19}  & *** & \myPosBar{0.20}\\
     \hline \hline
                     & {Adjusted R-squared} &  & ~0.84            &     & ~0.80            &     & ~0.70            &     & ~0.54\\ 
                     & {Moran's test}       &  & ~0.03            &     & ~0.02            &     & ~0.04            &     & ~0.05\\ 
	 \end{tabular}
	 \caption{Temporal analysis of Airbnb offering.}
    \label{tab:AirA_temporal}
\end{table*} 

\begin{description}
\item {\em 2012}. At early stages of adoption, the most important predictor is geography, i.e., Airbnb penetrated areas close to the city center first. Early adopters were likely young and ethnically-diverse residents living in central neighborhoods. A certain percentage of these early adopters might be composed of students, given the negative correlation with employment.

\item {\em 2013}. The coefficient for the variable distance from the center decreases in magnitude, and FourSquare check-ins stop being statistically significant, suggesting that, at a second stage, Airbnb  penetrated areas whose residents are not necessarily tech-savvy youngsters. Those residents tend to be of two types: the first tend to own their own houses, while the second tend to struggle financially (the income variable becomes negatively correlated). 

\item {\em 2014 and 2015}. The trends described for the year 2013 continue. In particular, the strongest predictors of Airbnb offering are two: low income and number of rented houses. Again, this indicates the possibility that Airbnb is helping people who might be struggling economically.
\end{description}

To sum up, we spell out three main insights. First,  central areas become consistently less predominant year after year: the coefficient \textit{distance} from the center decreases in magnitude, and the \textit{foursquare}  metric stops being statistically significant. Second, the correlation with \textit{income} becomes increasingly negative year after year -- late-adopting  hosts joined Airbnb for extra income. Finally, the correlation with \textit{owned} properties  becomes increasingly negative too --  late-adopting hosts did not tend to own their properties.

\subsection{RQ4. Where Do Airbnb Customers Actually Go?}
\label{sec:reviews}

We are aware that number of listings does not fully reflect the number of hosting events. Therefore, in this section, we study Airbnb demand by using the number of user reviews as a proxy. According to~\cite{fradkin2015bias}, the completion rate for reviews is high in Airbnb: the number of reviews over the number of stays is more than 70\%, making the number of reviews a good proxy for demand.

Using the same regression model of the previous analyses, we study how demand is associated with neighborhood socio-economic conditions over the past four years, from 2012 to 2015. In Section~\ref{sec:temporal-adoption}, when looking at the offering, we showed that Airbnb is first adopted in  central areas, but it then moves to more diverse areas of the city. By contrast, we do not observe such an evolution pattern for Airbnb demand. Instead, reviewing patterns year after year are very similar, if not constant.
Because we did not observe a temporal difference, we report the estimate for the year 2015 only  in Table~\ref{tab:AirR}: areas with high Airbnb demand are touristic (as one would have expected); they are close to the city center, and have high number of FourSquare check-ins and high population density.

\begin{figure}[t]
\centering
  \begin{subfigure}{.4\textwidth}
    \centering
    \includegraphics[width=0.98\linewidth]{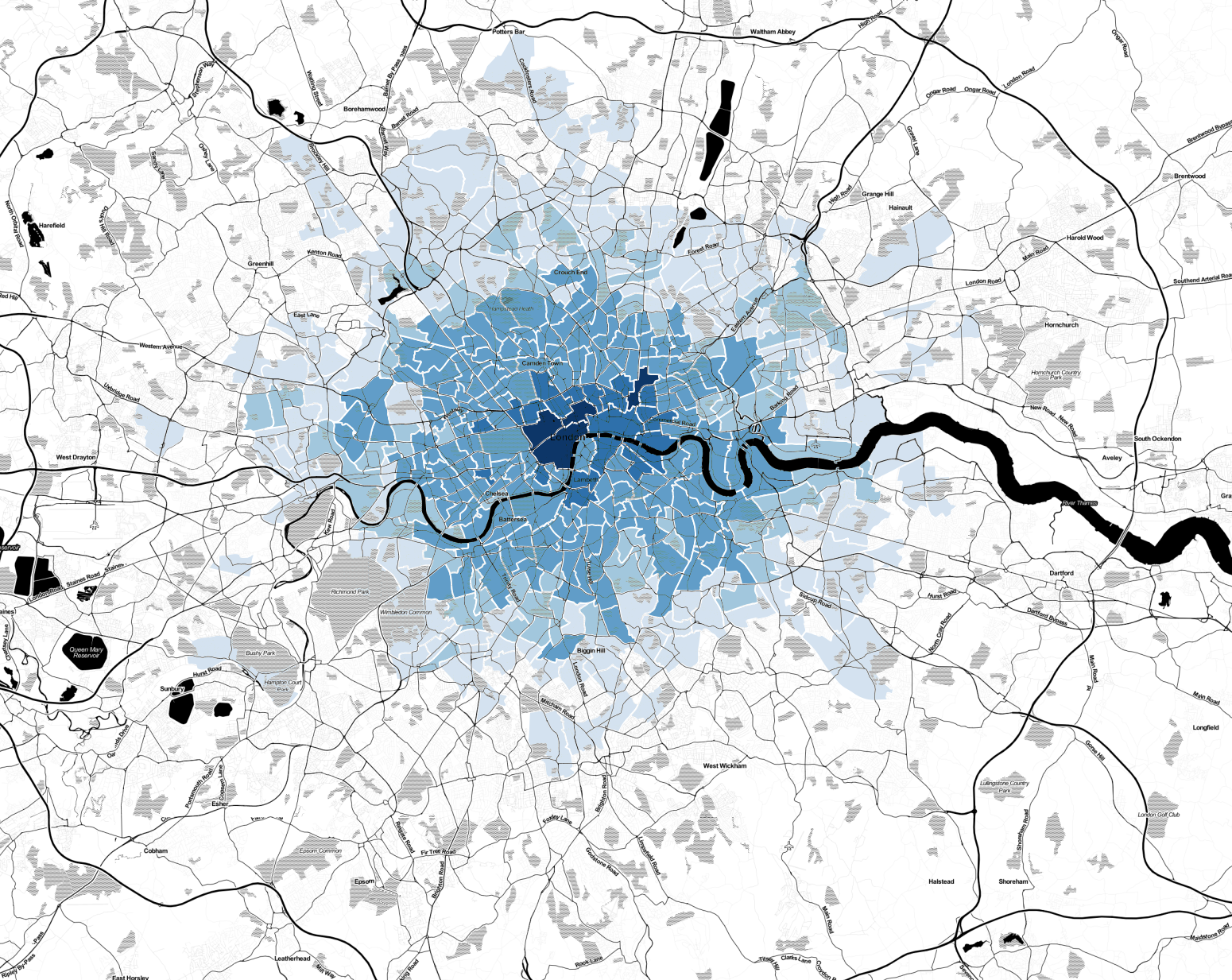}
  \end{subfigure}%
  \begin{subfigure}{.08\textwidth}
    \centering
    \includegraphics[width=0.98\linewidth]{figs/legend}
  \end{subfigure}%
  \caption{Heat map of the number of Airbnb reviews in each London ward. The darker the ward, the higher the number of reviews.}
  \label{fig:map-reviews}
\end{figure}

From the heat map of Figure~\ref{fig:map-reviews}, we confirm that areas of high Airbnb demand are closer to the center. Interestingly, by comparing the distribution of Airbnb rooms (Figure~\ref{fig:airbnb-rooms}) with that of Airbnb demand, we observe that Airbnb rooms cover a larger portion of the city. This indicates that many properties that are listed,  but are too far form touristic areas, are not being rented out. Therefore, while, in theory, Airbnb allows  travelers to be flexible when choosing the locations for their stays (which has the potential to distribute travelers among more diverse areas of the city), in practice, such flexibility is not fully exploited at the moment. 


\begin{table}[t]
    \mytablesize
    \tabcolsep 3pt
    \centering
	 \begin{tabular}{llrl}
     \hline \hline
                     & {\em ~~Indep. var}      & $p$-val & ~~$\beta$ \\
     \hline \hline
     Hotel           & $hotel\_\mathit{offering}$        &     & \myNegBar{-0.02}  \\
     Geography       & $distance$       & *** & \myNegBar{-0.18}  \\ 
     Attractiveness  & $foursquare$          & *** & \myPosBar{0.24}   \\ 
                     & $transport$     &     & \myNegBar{-0.07}   \\
                     & $attractions$               &     & \myPosBar{0.02}    \\
     Demographics    & $young$        & *** & \myPosBar{0.38}   \\ 
                     & $income$       & *   & \myNegBar{-0.12}  \\ 
                     & $employment$     &     & \myPosBar{0.07}    \\ 
                     & $ethnical\_mixed$ & **  & \myPosBar{0.12}   \\
                     & $bohemian$     &     & \myPosBar{0.00}   \\
                     & $melting\_pot$        & *** & \myNegBar{-0.13}   \\
                     & $education$    &     & \myPosBar{0.00}    \\
     Housing         & $living$  & .   & \myPosBar{0.07}    \\ 
                     & $green\_space$         &     & \myPosBar{0.05}   \\
                     & $top\_house\_price$      &     & \myPosBar{0.06}   \\ 
                     & $houses\_vs\_flats$         &     & \myNegBar{-0.09}   \\
                     & $owned\_vs\_rented$          & **  & \myNegBar{-0.22}   \\
                     & $house\_price$        & **  & \myPosBar{0.12}    \\ 
                     & $sold\_houses$         &     & \myPosBar{0.01}    \\
     \hline \hline
                     & {Adjusted R-squared} &  & ~0.85   \\ 
                     & {Moran's test}       &  & ~0.01   \\ 
	 \end{tabular}
	 \caption{Analysis of Airbnb demand.}
    \label{tab:AirR}
\end{table}


\mbox{ } \\

\section{Discussion}
\label{sec:discussion}

Based on our results, we now provide five main recommendations about how municipalities should set (Section~\ref{sec:disc:regulating}), enforce (Section~\ref{sec:disc:regulating}) and refine regulations (Section~\ref{sec:disc:refining}). We conclude this section by pointing out some limitations (Section~\ref{sec:disc:limitations}).

\subsection{Regulating}
\label{sec:disc:regulating}

To  properly regulate short-term rentals, a city needs to think about how, where, when, and what to regulate. 

\textbf{How.}  We envision a regulatory framework similar to that proposed by Stephen Miller in which short-term rental market is legalized through ``transferable sharing rights''~\cite{miller2015first}.  Each house owner has the right to  engage in a short-term rental for a given period of time.   To ensure market efficiency,  the transfer of rights needs to be effective, and entrepreneurs might be able to help. In fact, one way of ensuring effectiveness is to create web platforms  that sell transferable sharing rights in a way similar to what StubHub does when selling tickets~\cite{miller2015first}. Web platforms make it possible to  adjust prices  based on market demand in  real-time. That demand might be altered to some degree by municipal policy, not least because the rental terms would change depending on the city's tourism market. Since our analyses have shown that Airbnb has impacted different areas in very different ways (Section~\ref{sec:conditions}), a neighborhood's sharing rights might also be allocated depending on the  plan of the neighborhood's economic development. This right can be sold to others, if the owner does not wish to engage. The revenues generated by the sharing right market would  go to both the city council, which would be able  to raise revenues without raising taxes any further; and to neighborhood groups, which would be compensated for any externality. 

~

\fbox{\begin{minipage}{25em}
 \emph{Recommendation 1:} New web platforms should be built to offer schemes of ``transferable sharing rights''  in which prices are based on both real-time market demand and municipal policies. Policies should deal with the externalities created by the short-term rentals while capitalizing on the opportunities offered by them (e.g., decentralization of economic activity). Also, policies might be neighborhood dependent, in that, they might change across the neighborhoods of the same city. 
\end{minipage} }

~

\textbf{Where \& When.} It is important for municipalities to regulate where the permits get allocated because:
\begin{enumerate}
\item \emph{Initial conditions matter.} Based on our temporal analysis (Section~\ref{sec:temporal-adoption}), we have found that initial geographic conditions greatly influence which areas tend to benefit from the sharing economy in the end, and which do not. 

\item \emph{Local economies benefit.} Airbnb can be used as an economic development tool. It has been shown that Airbnb guests spend a considerable part of their money in the hosting communities~\cite{edelman2015efficiencies}. 

 \item \emph{Tourism should be sustainable}. One of the main priorities of local governments is to make tourism sustainable. In large cities, tourists tend to congregate in central areas, and residents often cannot cope with the increasing demand. Local governments are studying strategies for distributing tourism across the entire city. Our analysis has shown that, as opposed to hotels, Airbnb listings have a wider geographic coverage (Section~\ref{sec:overview}) and, consequently, naturally load balance tourists across the city. 
 
\item \emph{Concentration of short-term rentals has to be avoided.} If a neighborhood has an excessive number of short-term rentals, then its character and ambiance are bound to be compromised. Within the framework we are envisioning, municipalities should be able to limit the number of sharing rights.
\end{enumerate}

\fbox{\begin{minipage}{25em}
 \emph{Recommendation 2:} Transferable sharing rights should be allocated while considering four main factors: future consequences for adoption, development of local economies, sustainability of tourism, and avoidance of short-term rental ``hot-spots''.
\end{minipage} }

~

\textbf{What.} Sharing economy platforms are quite different from each other, and regulations should be tailored to each situation. The taxi industry and the hotel industry do not have the same legal framework; neither should Uber and Airbnb.  Additionally, as we have seen in the case of Airbnb for different categories of listings, important differences exist even within the same platform. It is therefore crucial to understand \emph{what} to regulate. Based our findings, we think that listings of rooms and houses should be regulated differently because:
\begin{enumerate}
\item  \emph{The socio-economic conditions are different.}  As opposed to houses, rooms tend to concentrate in low-income yet highly educated part of town (likely students) with a predominant non-UK born population (Section~\ref{sec:rooms-vs-houses}). Houses, instead, tend to be in wealthy areas.

\item \emph{The social consequences are different.} Central neighborhoods are increasingly becoming places in which  properties are rented by wealthy people (Section~\ref{sec:conditions}). As a consequence, in the long term, the social fabric of those neighborhoods is likely to be compromised, if the situation is left unregulated. Studies have shown that it takes time (years) to build what Putnam calls ``social capital'' among neighbors~\cite{putnam2001bowling}, and having a critical mass of short-term renters does not help. Also, happiness might be affected, as a good predictor of it is the number of people one personally knows and regularly meets in his/her neighborhood~\cite{layard05happiness}.

\end{enumerate}

\fbox{\begin{minipage}{25em}
 \emph{Recommendation 3:}  The terms of transferable sharing rights should change depending on whether a room or an entire apartment is rented. 
\end{minipage} }

~

\subsection{Enforcing}
\label{sec:disc:enforcing}
Regulations are effective only if they are enforced. An important part of such an enforcement is to be able to identify offenders. One way of doing so is to automatically spot anomalous behavior from data, as the retail banking usually  does. By matching Airbnb data with census data, we have been able to find out that Airbnb rooms tend to be offered disproportionately in areas where people rent (Section~\ref{sec:temporal-adoption}). In London, this means that tenants engaging in such short-term letting almost certainly violated general ``rental agreements'' on subletting.  One could easily build an index of ``subletting violation'' by cross-correlating the two data sources of Airbnb rentals and of house ownership.   However, this would be possible only if municipalities incentivize the creation of a data sharing ecosystem. Sharing economy companies can and should share part of their data too. This data should be sufficiently specific to inform policies, but also fairly vague to protect the privacy and safety of customers. 

~

\fbox{\begin{minipage}{25em}
 \emph{Recommendation 4:}  Municipalities should incentivize the creation of a data sharing ecosystem. 
\end{minipage} }

~

\subsection{Refining}
\label{sec:disc:refining}

After defining and enforcing regulations, a city needs to engage in a dialog with citizens. Sharing economy platforms could provide data upon which the city evaluates the impact of the short-term rental market (e.g., the increasing demand on public services) and  refines its  responses to it. 

~

\fbox{\begin{minipage}{25em}
 \emph{Recommendation 5:}  Municipalities should constantly evaluate the impact of short-term rentals based on data, and they should accordingly refine their regulations.  
\end{minipage} }

~

\subsection{Limitations} 
\label{sec:disc:limitations}
Our study has two main limitations. The first is that the analysis is limited to the city of London. Therefore,  generalization of our results to other cities might be inappropriate, as both Airbnb adoption and socio-economics characteristics are very heterogeneous across cities. Second, while we do have longitudinal data for Airbnb, and therefore we observe temporal and geographical variation of its adoption over time, we only have cross-sectional data for the socio-economic metrics.  This makes it difficult to study causal mechanisms. In the future, to partly address that issue, we plan to extend our study to a variety of cities by resorting to the Airbnb data made freely available on \url{http://insideairbnb.com/}, and by further collecting longitudinal socio-economic data.


\section{Conclusion}
Only a few efforts (isolated cases, e.g., Portland, Oregon)\footnote{See: \small\url{http://www.oregonlive.com/front-porch/index.ssf/2014/07/portland_legalizes_airbnb-styl.html}} have been devoted to the regulation of the sharing economy, and, even in those cases,  hard-and-fast rules have been laid out. 
By contrast, this work has called for evidence-informed policy making. Cities should rely on data analysis to envision and revise their local ordinances, and here we have shown the way by analyzing data about short-term rentals to offer regulatory recommendations.

We have used London as a living lab. We have studied data collected unobtrusively on how  Airbnb has turned out to be in a fairly unregulated context.  A lot of the demand for short-term rentals comes from touristic areas. Those areas change over time, and so traditional regulations are unlikely to be able to respond to those changes. That is why,  based on our findings, we have drafted five main recommendations for regulating Airbnb. 

Our attempt contributes to the general idea of ``algorithmic regulation'', which argues for the analysis of large sets of data to produce regulations that are responsive to real-time demands. Such an approach might be used to regulate any civic issue independent of the sharing economy.

Future work should propose comprehensive evidence-informed legal frameworks, thanks to which a city is able to welcome both the sharing economy and visitors  from all over the world, while still feeling home to its residents.




\section*{Acknowledgement}

We thank Stephen Miller for his extremely helpful feedback. 

\bibliographystyle{abbrv}
\bibliography{references}  

\end{document}